\documentstyle[11pt,newpasp,twoside,epsf]{article}
\def\edcomment#1{\iffalse\marginpar{\raggedright\sl#1\/}\else\relax\fi}
\reversemarginpar

\begin{document}
\title{Effects of Fluid Instabilities on Accretion Disk Spectra}
\author{Shane Davis, Omer Blaes, Neal Turner and Aristotle Socrates}
\affil{Department of Physics, University of California, Santa Barbara, CA  93106}

\begin{abstract}

Numerical calculations and linear theory of radiation 
magnetohydrodynamic flows indicate that the photon bubble 
and magnetorotational instability (MRI) may produce 
large density inhomogeneities in radiation pressure supported media. 
We study the effects of the photon bubble instability on 
accretion disk spectra using  2-D Monte Carlo (MC) and 1-D Feautrier 
radiative transfer calculations on a snapshot of a 2-D numerical 
simulation domain. We find an enhancement in the thermalization 
of the MC spectra over that of
the Feautrier calculation.  In the inner-most regions of these 
disks, the turbulent 
magnetic pressure may greatly exceed that of the gas.  It is then
possible for bulk turbulent Alfv\'enic motions driven by the MRI to exceed 
the thermal velocity making turbulent
Comptonization the dominant radiative process. We estimate the spectral
distortion due to turbulent Comptonization utilizing a 1-D MC
calculation.

\end{abstract}

\section{Introduction}

The standard assumption central to most accretion disk models is that the 
turbulence responsible for angular momentum transport and gravitational 
energy release can be parameterized by an ``$\alpha$-viscosity'' prescription 
(Shakura \& Sunyaev 1973). The hypothesis that turbulence drives accretion
has been borne out by the  identification of the magnetorotational 
instability (MRI, Balbus \& Hawley 1991) as a generic source for
magnetohydrodynamic turbulence in differentially rotating flows.
However, it is not yet known how accurately $\alpha$-disk
models describe real accretion disks. It is possible that
both the structure and emitted spectrum of a disk in which the MRI is active
may deviate significantly from the predictions of ad hoc  $\alpha$-disk 
models. 

Numerical simulations of the radiation magnetohydrodynamic (RMHD) equations 
(Turner et al. 2003) show that 
MRI turbulence can produce a factor of 20 variation in density. In addition
to the MRI, local linear analyses of the RMHD equations demonstrate that a 
radiatively driven photon bubble instability exists in radiation 
dominated accretion disks (Gammie 1998). This instability is present as 
long as the 
magnetic field is finite, but the growth rate for the instability is 
greatest when the magnetic pressure exceeds the gas pressure (Blaes \& Socrates 
2003). Numerical RMHD calculations of radiation pressure supported disks 
which neglect
shear (Turner et al., in preparation) show that the non-linear 
development of the photon bubble instability does indeed produce large density 
variations.  In section 2 we discuss the effects of these density
inhomogeneities on the emergent spectrum.

Fluid motions associated with the MRI may also influence the emission.
MRI turbulence is fundamentally magnetic in nature with a characteristic 
velocity $v_w$ given by the Alfv\'en speed $v_A$ which is less than the local 
sound speed $c_s$.  Close to the hole, where the radiation pressure is 
dominant, $c_s$ is given by the radiation sound speed which is larger than 
the sound speed of the gas.  Thus, it is possible for the MRI turbulence to be 
{\it highly supersonic relative to the gas} and 
Compton scattering off of the turbulent eddies may produce a larger spectral
modification than thermal Comptonization (Socrates, Davis,
\& Blaes 2003, hereafter SDB03).  In section 3, we summarize these results.

\section{Monte Carlo Results}

We calculate a full-disk-thickness numerical solution of the RMHD equations in
a stratified, radiation supported disk using the ZEUS-2D code with flux 
limited diffusion. Rapid photon diffusion creates large density
inhomogeneities with relatively small temperature gradients. We then
perform 2-D MC and 1-D Feautrier radiative transfer calculations through 
the uppermost quarter of the simulation domain. We horizontally average the 
domain to create a 1-D density and temperature input profile for the Feautrier 
calculation.  Both methods model the effects of electron scattering and 
free-free emission and absorption. The optical depth at the base of the MC 
calculation was chosen to be several times the unit effective optical depth.

In figure 1, we compare the emergent spectrum from the MC calculation 
with the Feautrier spectrum and blackbody emission at a representative 
domain surface temperature.  Though both of the calculated spectra resemble
modified blackbody curves, the 2-D calculation's spectrum is more thermal 
(closer to blackbody) than the 1-D result. The stronger density dependence of 
absorption and emission relative to scattering is responsible for this
difference.
In the inhomogeneous 2-D MC domain, the surface flux is largest in the low
density regions but is dominated by photons
{\it emitted} in the densest regions where absorption is
increasingly probable relative to scattering.  Thus, on average, the photons 
escaping through the surface in the 2-D model will undergo fewer 
scatterings than those photons reaching the surface in 
the 1-D model resulting in a ``less-modified'' blackbody spectrum.
This enhanced flux exceeds the local Eddington limit by 
a factor of order ten, consistent with predictions for highly inhomogeneous
atmospheres (Begelman 2001).

\begin{figure}[t]
\plotone{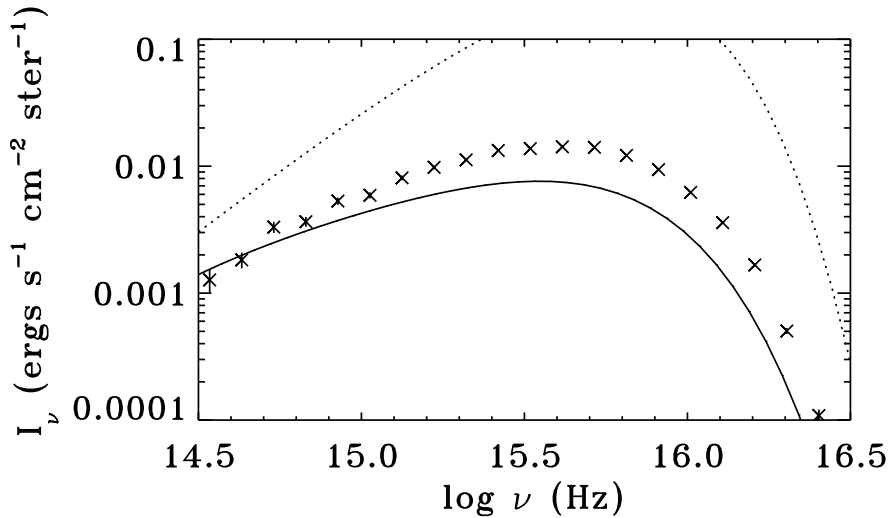}
\caption{The output of 2-D MC calculations (x's) and
1-D Feautrier method (solid line) are compared with a 
$1.06 \times 10^5 \rm K$ black body (dotted line).}
\end{figure}

\section{Turbulent Comptonization}

In order for turbulent Comptonization to be important in accretion disks,
the bulk velocity of the turbulent eddies must exceed the thermal velocity
of the electrons. To compare the turbulent and thermal velocities, it is 
convenient to define a turbulent wave temperature $T_w$ in terms of the 
turbulent velocity on the outer scale. The wave temperature can then be
related to thin disk parameters (SDB03)
 
\begin{equation}
T_w \equiv \frac{m_e}{3 K_B}\left<v^2_w\right> \simeq 9.0\times 10^{9}\,\frac{\alpha}{\epsilon^2}\left(\frac{L}{L_{edd}}\right)^2\,r^{-3}\, {\rm K}
\end{equation}
 
Here, L is bolometric luminosity,  $\rm L_{edd}$ is the Eddington
luminosity, $\epsilon$ is the radiative efficiency, $r$ is 
the cylindrical radius in units of gravitational radii $r_g=GM/c^2$, 
$\alpha$ is the standard viscosity parameter, and the effects of general
relativity and no-torque inner boundary conditions have been neglected. 
It can be seen from 
eq. (1) that the wave temperature dominates the thermal temperature 
near the inner radius of the disk for sufficiently high accretion rates. 
The photons can only sample the velocity field on scales of order the 
photon mean free path such that the relevant wave temperature 
is generally less than that given by eq. (1). Our lack
of understanding of turbulent dissipation in accretion disks hinders our
ability to model their vertical structure. This 
uncertainty is further
compounded by the possibility that turbulent Comptonization may itself be 
a dominant means of energy release.
Therefore, we adopt a simple
homogeneous model to estimate the reduction in wave temperature appropriate
for the reduced velocities.

1-D MC calculations show that for sufficiently high turbulent
stresses, turbulent Comptonization may account for 
a significant amount of the far-UV and X-ray power of accreting sources.
  For example, 
the spectrum shown in figure 2 has significant power up to $\sim$1 keV 
and resembles soft X-ray excesses observed in many Seyfert 1 galaxies
and QSO's. If soft excesses in these sources are due to turbulent
Comptonization, they provide a direct means for probing the turbulence.

\begin{figure}[t]
\plotone{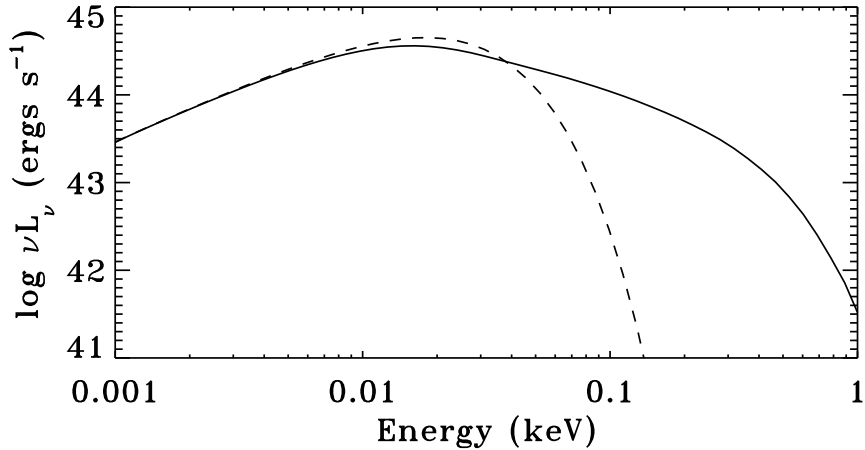}
\caption{A spectral energy density produced by 1-D MC calculations for 
an accretion disk around a $10^8 M_\odot$ maximally spinning Kerr 
black hole. The dotted line is the thermal disk spectrum neglecting the
effects of turbulent Comptonization.}
\end{figure}
\section{Conclusions}

Large density fluctuations due to the photon bubbles instability may alter
the spectrum emitted in radiation pressure dominated accretion disks. These 
results are based on
calculations which neglect shear.  If shear 
is not neglected the MRI may affect the photon bubble 
instability, possibly altering the disk structure. However, for sufficiently 
large $v_A$, the MRI may be compressible in radiation dominated regions 
and may produce significant inhomogeneities on its own (Turner et al. 2003). 

We also analyze the interactions of photons with turbulent eddies and show 
that for sufficiently large turbulent stresses, a 
prominent spectral component in the far UV and X-ray bands emerges.  
Observations of
accreting systems may provide a means to directly probe the mechanism 
responsible for angular momentum transport in these sources.

\end{document}